\documentclass[12pt,twocolumn]{article}
\usepackage{fixltx2e}
\usepackage{graphicx}
\usepackage{paralist}
\author{Travis Denny\\
Department of Computer Science\\
Oklahoma State University, Stillwater, OK 74075, USA}
\title{Faked states attack and quantum cryptography protocols}
\date{\today}
\begin{document}
\maketitle
\begin{abstract}
Leveraging quantum mechanics, cryptographers have devised provably secure key sharing protocols\cite{bennett:84,bennett:92,ekert:91,scarani:04}. Despite proving the security in theory, real-world application falls short of the ideal. Last year, cryptanalysts completed an experiment demonstrating a successful eavesdropping attack on commercial quantum key distribution (QKD) systems\cite{lyderson:10}. This attack exploits a weakness in the typical real-world implementation of quantum cryptosystems. Cryptanalysts have successfully attacked several protocols\cite{gerhardt:11}. In this paper, we examine the Kak quantum cryptography protocol\cite{kak:06} how it may perform under such attacks.
\end{abstract}
\section{Introduction}
Two parties, Alice and Bob, wish to discuss some rather sensitive information. They want to have a private conversation. A third party, Eve, wishes to listen in on this private conversation. Since Eve's propensity for eavesdropping is well-known, Alice and Bob try to design a system that provides secure communication. Since Eve is the determined sort, she will search for and exploit any weaknesses in the system created by Alice and Bob. Alice and Bob will eventually discover their system leaks (Eve is a notorious gossip) and they will either try to patch it or create a new system. Escalation ensues, and being three clever people, increasingly clever methods are devised for secure communication and for cracking those systems, or, cryptography and cryptanalysis, respectively.

So far, the only clever cryptographic method that cryptanalysts have not cracked is the one-time pad. The one-time pad creates a ciphertext using a key at least as long as the plaintext input and the key is used only once. If Alice and Bob both possess the same secret key, Alice can encipher the plaintext, send the result to Bob, and he can decipher with his copy of the key. The one-time pad has been proven perfectly secure\cite{shannon:49}, but it suffers from some non-trivial requirements. Namely, perfectly secure system needs a true random number generator and a secure way to distribute the key. Current cryptographic random number generators remain unproven. Distributing the key appears to provide the exact same challenge cryptographers were trying to address in the first place: sending a secure message. It turns out that both of these requirements may be satisfied by leveraging quantum mechanics.
\section{Quantum Cryptography}
Approaching the two requirements for a one-time pad with quantum mechanics returns promising results. First, using quantum mechanics to generate random numbers seems ideal\cite{bell:64}. In fact, commercial quantum random number generators are available\cite{idQ}. Second, one can imagine using quantum mechanics to set up a secure short-term communication channel for distributing the key. One quantum key distribution (QKD) protocol serves as a foundation for nearly all of the rest. The Bennett-Brassard 1984 (BB84) relies on two principles of quantum mechanics:
\begin{enumerate}
\item One cannot take a measurement without disturbing the system.
\item One cannot clone an unknown quantum state.
\end{enumerate}
Under this protocol, Alice will send quantum bits (qubits) to Bob in the form of photons transmit the key. Should Eve intercept those photons, she will not be able to ``read'' the key without alerting Alice and Bob to her presence. A closer look at the basics of the protocol will illustrate how this is true.

\begin{figure*}[tbp] \centering
\includegraphics[width=350px]{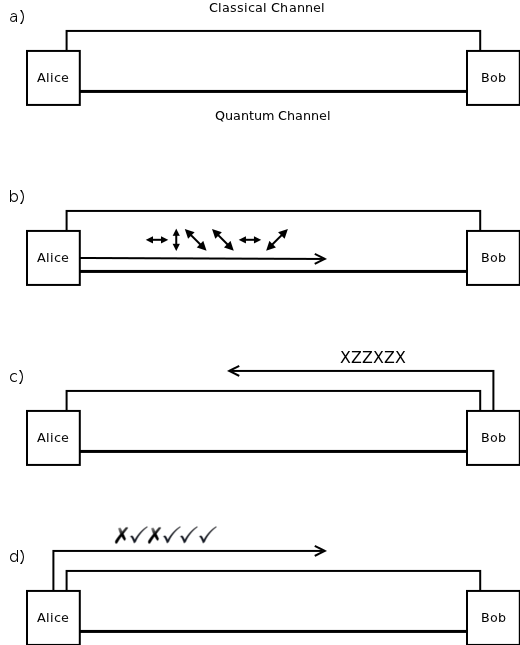}
\caption[BB84 Protocol]{
\begin{inparaenum}[\itshape a\upshape)]
\item Basic setup between Alice and Bob. The classical channel is insecure and it is a non-issue for Eve to eavesdrop on communication over this line. The quantum channel is where the key is transmitted.
\item Alice sends photons polarized in such a way that if they are measured in its complementing base, it will produce a $1$ or (exlusive or) $0$ reliably.
\item Bob measures the photons with his random basis, and sends the basis over the classical line to Alice.
\item Alice tells Bob which elements of his basis were compatible with her original signal. The qubits that Bob measured ``correctly'' make up the \emph{sifted-key}.
\end{inparaenum}
\label{bennetbrassard}
}
\end{figure*}

This protocol uses polarization encoding.
Photons are prepared in four types of polarizations: horizontal, vertical, and two diagonals. Figure~\ref{pe} illustrates these four encodings. Actually, there is another polarization, circular, but we will come back to that. Say vertical and $-45\,^{\circ}$ diagonal polarization represent 0, and horizontal and $+45\,^{\circ}$ diagonal polarization represent $1$. To measure a qubit, Alice and Bob will use one of two bases. One base, X, measures diagonally polarized photons. The other, Z, measures horizontally and vertically polarized photons. Figure~\ref{bennetbrassard} shows how the BB84 protocol works. Now, that is not to say one cannot measure horizontally polarized photons in the X base. In fact, measuring with the “wrong” base reveals the beauty of this protocol.
\begin{table}[hbp]
\begin{tabular}{|c|c|}
\hline
\multicolumn{2}{|c|}{Polarization Encoding}\\
\hline
\includegraphics{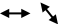} & $0$ \\
\hline
\includegraphics{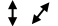} & $1$ \\
\hline
\end{tabular}
\caption{Typical configuration for polarization encoding.\label{pe}}
\end{table}

The first principle of quantum mechanics, mentioned above, says that measuring disturbs the system. This means that the quantum state collapses to a classical state upon measurement. For example, if one measures a horizontally polarized photon in the Z base, the measurement will return a 1. If one measures a horizontally polarized photon in the X base, the measurement will return a 0 or a 1 with equal probability. The quantum state collapses to either $-45\,^{\circ}$ or $+45\,^{\circ}$ diagonal polarization. When measuring in the Z base, the quantum state still collapses, but it collapses to horizontally polarized every time. The results are the same for all four of the polarizations. Therefore, measuring in the ``right” base produces one of the two results (1 or 0) every time, while measuring in the ``wrong” base produces one of the two results with 50\% probability. Alice will send a series of qubits so Bob can measure with a series of bases, called a basis, and get a string of 1s and 0s.

How is this useful? If Alice wants to send a key via photons to Bob, she will have to tell him, over classical communication lines, which basis to measure with. Otherwise, 50\% of the 1s and 0s Bob possesses at the end will not match Alice's bitstring. If Alice communicates the basis to Bob over classical lines, it will be a non-issue for Eve to eavesdrop on the classical line, intercept the qubits, measure in Alice's basis, and generate and send qubits to Bob with the same polarization as Alice's original signal. Eve will have the same key as Alice and Bob and can eavesdrop on their private conversation.

Alice does not send her basis, so Bob measures in a randomly chosen basis. This results in 50\% error in Bob's key. After Bob does his measurement, he sends \emph{his} basis to Alice. Alice compares her basis his and tells Bob which elements of his basis match hers. Alice and Bob both discard the mismatches in their results, producing a common \emph{sifted key}.

Eve cannot perform the same attack she used when Alice sent the basis measurement to Bob. Alice's qubits have come and gone, so there is no opportunity to measure in Bob's basis unless Eve can discover it in advance. This leaves Eve with no choice but to measure the qubits in her own random basis. However, 50\% of Eve's basis does not match Alice's basis. Eve does not know which elements of her basis are ``wrong”, but she must send something on to Bob or he and Alice will simply discard the missing qubit bringing Eve no closer to her goal. The best Eve can do is send a guess.

If Eve guesses the base to send to Bob, half of those will be incongruent with Alice's original qubits. Alice and Bob will only use about half of the qubits. Now, half of those sifted qubits will produce unmatching measurements for Alice and Bob (ie. Eve has introduced a 25\% error rate in the sifted key between Alice and Bob). The error rate in the sifted key will alert Alice and Bob to Eve's presence, and they can discard the key ensuring Eve gains no useful information.

Why can Eve not simply copy the qubit to send to Bob? Then, she could wait until Bob announces his basis and perform the same measurement. Going back to the second featured principle of quantum mechanics this protocol relies on, one sees that cloning unknown quantum states is impossible\cite{wootters:84}. Since Eve cannot clone the qubit, her only option that remains is sending a guess.

Theoretically, this protocol is provably secure.
However, the real-world implementation proves less than ideal.
Furthermore, quantum operations suffer from errors like classical operations.\cite{kak:96,kak:99}
\section{Faked-state Attack}
The BB84 protocol relies on the system's ability to send and detect single photons. Sending single photons at an exact time is a problem\cite{bell:64}, so current systems send a very weak light signal instead. These signals are not quite single photons. Likewise, the detectors emulate detecting single photons by picking up very weak signals. It is this reliance on weak signals that eavesdroppers will attack. In the faked-state attack, Eve manipulates this weakness in Bob's detectors to force him to measure in the same basis as Eve\cite{makarov:05}.

The faked-state attack exploits a weakness in the apparatus of QKD systems, namely, the diode used to detect photons\cite{makarov:06,makarov:09}. The avalanche photodiode (APD) detector is supposed to detect single photons, or at least approximate the detection of single photons. However, an APD requires a recharge time of roughly $1\mu$s before it is ready to detect another photon. This normally would not be a problem because under expected use, weak laser signals would be sent infrequently enough for the detector to recharge. Makarov discovered that if you shine continuous light into the APD, the detector does not recharge and is reduced to a classical photodiode. Furthermore, by manipulating his continuous beam of light, he can make the detector click when he wants or he can blind the detector from seeing valid input.

Now, the apparatus has four detectors, one for each polarization. Eve will intercept the photon from Alice and prepare a faked-state to send to Bob. They say faked because Eve is not sending Bob a quantum state, but she is making Bob's apparatus think it is detecting a quantum state. Anyway, Eve will intercept Alice's photon, and she will prepare her faked state in the opposite base with the opposite bit she detects. For example, if Eve detects a 0 in the X base, she will prepare a 1 in the Z base for Bob. Here, she exploits the detectors. At the same time she sends her prepared fake state, Eve will blind Bob's 1-bit detectors. If he measures in the same base as Eve, X, he has a 50\% chance of detecting a 0 or nothing at all. If he measures in a base different than Eve's, Z, he is guaranteed to detect nothing at all. Figure~\ref{fsa} illustrates how this works given Alice's choice of sending 0 in the X base. Doing this, Eve guarantees Bob's apparatus only detects the same bit, in the same base, she detected.
\begin{table*}[tbp] \centering
\begin{tabular}{|c c c c c c|}
\hline
Alice & Eve's Base & Eve's Measurement & Eve Sends & Bob's Base & Bob's Measurement\\
X0 & X & 0 & Z1 & X & 0 or no detection\\
X0 & X & 0 & Z1 & Z & no detection\\
X0 & Z & 0 & X1 & X & no detection\\
X0 & Z & 0 & X1 & Z & 0 or no detection\\
X0 & Z & 1 & X0 & X & no detection\\
X0 & Z & 1 & X0 & Z & 1 or no detection\\
\hline
\end{tabular}
\caption[Eve's faked-stage attack]{Eve's faked-stage attack. Note Eve's measurement in the bottom for has a 50\% chance of detecting a 0 or a 1 since she uses an incompatible base to Alice's base.\label{fsa}}
\end{table*}

One might think Bob not detecting 50\% of Alice's signals would raise concerns, but under normal conditions with today's technology, Bob's apparatus will only detect a small portion of the photons sent by Alice. The blame for this lies with the system's reliance on psuedo-qubits, and this reliance is a result of the protocol's requirement that the system send and register single photons. However, not all QKD protocols have this requirement.

\section{Three-Stage Protocol}
Imagine Alice wants to send a secret item to Bob. She puts the item in a box and locks the box with her padlock. Then, she sends the box to Bob. When it arrives, Bob puts his own padlock on the box. Now, the box has two padlocks on it. Then, he sends it back to Alice. Alice unlocks her padlock and sends it back. Upon arrival this time, the box only has Bob's padlock on it. He unlocks it and withdraws the secret item. Assuming these padlocks are indestructible and can only be unlocked with their respective owners' key, Eve cannot obtain the secret item.

\begin{figure*}[tbp] \centering
\includegraphics[width=400px]{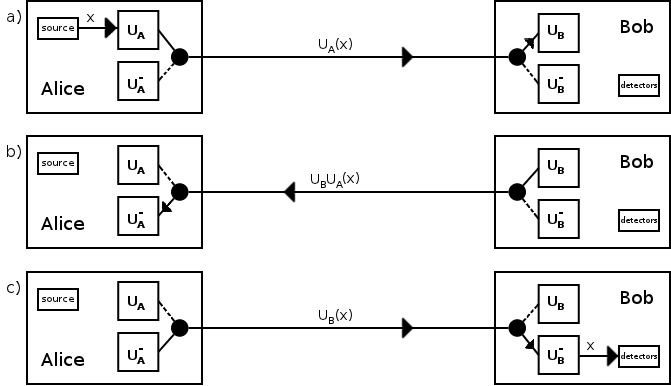}
\caption[Kak's Three-stage Protocol]{
\begin{inparaenum}[\itshape a\upshape)]
\item A qubit is emitted from the source, and Alice's transformation is applied. The qubit travels to Bob, where Bob's transformation is applied.
\item The qubit sent back to Alice, now the result of two transformations. Alice's inverse transformation is applied canceling out her initial transformation.
\item The qubit is sent to Bob one last time, now with only his transformation applied. Bob's inverse transformation is applied, and Bob's apparatus measures the qubit.
\end{inparaenum}
\label{3stage}
}
\end{figure*}

In the three-stage protocol\cite{kak:06}, the secret item is the symmetric key Alice and Bob want to use for a one-time pad.
The padlocks are transformations\cite{kak:00} applied to the qubits.
It is important that these transformations are commutative (ie. for transformations U and V, UV = VU).
Alice will perform her transformation (U\textsubscript{A}) on the qubit and sends it to Bob.
Bob applies his transformation (U\textsubscript{B}) and sends it back to Alice.
Alice performs her inverse transformation (U\textsubscript{A}\textsuperscript{$-$}) and sends it back to Bob.
Bob applies his inverse transformation (U\textsubscript{B}\textsuperscript{$-$}) and measures the qubit in the predetermined base to get the bit.
Figure~\ref{3stage} illustrates this transaction.

Unlike BB84, the three-stage protocol does not rely on Alice and Bob sending single photons.
The three-stage protocol provides another important advantage. %Lacks another weakness found in BB84
When she attacks communication under the BB84, Eve knows that the qubits will arrive in one of two polarization bases.
Then, she must only discern one bit of information.
In the three-stage protcol, qubits can be sent with \emph{any} polarization.
They are limited only by the precision of the equipment.

For example, let Alice and Bob use rotation for their transformations.
Their systems perform one of 1024 possible rotations and the appropriate inverse rotation, respectively.
Eve can easily cut into the line\cite{makarov:05} and siphon off some of the photons in any stage and allow the rest to continue to their intended destination.
Then, she can send these siphoned photons through a series of filters to discover the angle.
If Eve can siphon off and discover the angle at all three stages of the protocol, she can determine Alice's and Bob's transformations.
However, note that to determine the angle with 100\% certainty, Eve would have to siphon off at least 1024 photons (assuming she has the technology to split and direct them to 1024 different filters).
If Alice and Bob send less photons than this threshold, Eve's attempt will ``use up'' the photons, and she will be detected.

Surely, Bob has to use a detector similar to the APD to perform the final measurement.
Can Eve manipulate a beam of light to create a faked state within Bob's apparatus such that she can learn the key?
Assume Eve can create faked states within Bob's equipment.
She make his equipment register a 1, a 0, or nothing at all.
Eve cannot leverage this in any way to gain secret information.
Notice that intercepting the signal at any stage, Eve will possess a transformed qubit.
Measuring the transformed qubit has a 50\% chance of coinciding with Alice's original input.
She can force Bob's apparatus to register the same result, but the 50\% error rate will betray her eavesdropping.

Eve cannot leverage the faked states attack on the three-stage protocol, but she can employ a simpler intercept-resend attack.
Since it is safe to assume that Eve can obtain the same kind of equipment as Alice and Bob, she could cut into the line, pose as Bob and complete a transaction with Alice, and then pose as Alice and complete a transaction with Bob.
Now, she is privy to any secret communication over the quantum channel.

One possible solution to this attack is to apply classical cryptography\cite{chen:09} to ensure the message's authenticity.
Another solution uses trusted certificates created using quantum mechanics\cite{perkins:06}.

\section{Conclusion}
Real-world implementations of quantum cryptography protocols are not as completely secure as one may hope when looking at the theory.
Simply put, the fault lies with the apparatus.
In a more abstract sense, the fault lies with trying to emulate quantum states.
These emulations are imperfect and attackers can exploit the imperfections.
The Kak protocol offers advantages, but it is yet to be seen if it can be implemented for real-world applications and how it holds up against other attacks.
\bibliography{dennyQC}

\end{document}